\documentstyle[lscape]{mn}

\begin{document}

\title[Red Giant Branch Photometry]{The Analysis of Red Giant Branch Photometry in Galaxies}

\author[C. M. Frayn and G. F. Gilmore]
       {C. M. Frayn$^{1}$, G. F. Gilmore$^{1}$\\
       $^{1}$ Institute of Astronomy, University of Cambridge, Cambridge, CB3 0HA, UK\\
}
\date{Submitted to MNRAS, June 2002}

\pagerange{\pageref{firstpage}--\pageref{lastpage}}
\pubyear{2002}

\maketitle

\label{firstpage}

\begin{abstract}

We present an analysis of the many possible methods for simulating and
analysing colour-magnitude diagrams, with application to studies of the
field halo stellar populations of resolved galaxies. Special consideration
is made to the analysis of stars on the Red Giant branch (RGB), and methods
for obtaining metallicity distributions for old populations based on analysis
of evolved sections of the Colour Magnitude Diagram (CMD).  These tools are
designed to provide a reliable and accurate method for the analysis of
resolved Population II stars in the haloes of nearby galaxies.

In particular, we introduce a Perpendicular Distance method for calculating
most likely source isochrones for stars in our input dataset. This method is
shown to have several advantages over a more traditional approach
considering the isochrone as a set of finely-spaced Gaussian probability
distributions.  We also consider methods by which the obtained metallicity
distribution might be most efficiently optimised and especially evaluate the
difficulties involved in avoiding sub-optimal local maxima in the likelihood
maximisation procedure.

\end{abstract}


\section{Introduction}

Stellar populations in galaxies encode the star formation and merger
histories of galaxies, the relationship between gas in- and out-flows and
the local star formation rate over time. Extensive studies of the stellar
populations in the central regions of galaxies, bulges and disks, have
quantified knowledge of the dominant metal-rich populations. Galaxies with
substantial disks cannot have experienced recent major mergers -- thus their
field halo stars uniquely record their early history and minor mergers.

The stellar haloes of galaxies contain both the very oldest metal poor stars
(Population II) and the debris of continuing accretion, which may have a
wide range of ages and abundances. The relative importances of early
formation and later accretion are a key test of hierarchical galaxy
formation models, yet remain extremely poorly known.

In particular, it would be interesting to analyse nearby galaxies for
comparison with current theory. Simulations of hierarchical Cold Dark Matter
(CDM) cosmologies predict that massive ellipticals form at redshift $<1$
(Kauffmann 1996; Baugh, Cole \& Frenk 1996; Baugh et al. 1998).  This may
disagree with many current observational results (e.g. Daddi, Cimatti \&
Renzini 2000), who suggest that many large ellipticals were already in place
well before this. It is an important task, therefore, to analyse stellar
populations in a robust and realistic manner.

A task of fundamental importance in this problem is determining the
metallicity distribution function as this distinguishes between simple
coeval populations, and those accumulated over a longer time period in a
continually enriched interstellar medium.  This would allow us to
discriminate the roles of infall, accretion and recycling of materials on
galactic time-scales.  Most importantly it would allow us to distinguish
between an initial burst of unenriched star formation at high redshift and
later bursts triggered by the merger of gas-rich clumps of higher abundance.
Do protogalaxies lose their gas before merging, or does substantial star
formation still occur during the merger process in the sense proposed for
globular clusters by Ashman \& Zepf (1992), and observed in, for example,
the Antennae galaxies by Whitmore \& Schweizer (1995) and Whitmore et al.
(1999)?

In this paper we discuss a set of tools developed for the analysis and
interpretation of the halo stellar populations in nearby galaxies.  We
discuss firstly the method by which artificial stellar populations may be
produced in order to compare with real datasets.  In the third section, we
discuss the way we can build on this work in order to fit metallicity
distributions to stellar populations, given accurate photometry of their RGB
populations.  We explicitly include practical algorithm implementation
details, and describe several alternative approaches, as such a description
is crucial to applications in astrophysics. We apply these techniques to HST
data in forthcoming papers.

\section{Creating artificial colour-magnitude diagrams}

The method of stellar population synthesis has been used since the early
1980s (see eg Bruzual, 1983) primarily for the analysis of unresolved
stellar populations.  It was refined by Charlot \& Bruzual (1991) and later
by Bruzual \& Charlot (1993).  They proposed a new method of populating
colour-magnitude space by interpolating between a large library of
accurately calculated theoretical isochrones.  This is the method adopted by
this work, here applied to resolved populations.

The Bruzual \& Charlot method relies on the fact that any stellar population
can be considered as the linear sum of single stellar populations of
delta-function-like distribution in age and metallicity (Z).  This holds
true if the spacing between these constituent functions is small enough to
be insignificant compared to the observational errors.

To generate a complex stellar population, therefore, requires a simple
linear sum of a set of appropriately normalised single stellar populations
each generated from one single isochrone.

  \subsection{Isochrone selection}

The isochrones used in this study are those by Girardi et al. (2000).  We
shall consider only $V$ and $I$ CMDs here, although the methods described
apply equally well to any suitable choice of filters.

Uncertainties in the choice of the appropriate stellar model can result in a
very different isochrone for some stages of stellar evolution.  See, for
example, the work done on convective overshooting by Bertelli et al. (1990)
and that on alpha-enhancement by Salaris, Chieffi \& Straniero (1993).
Isochrone matching problems are considered in detail in our second paper (in
press), together with the intrinsic systematic errors inherent in the
technique of isochrone fitting.  We consider here the numerical techniques
required to implement the methodology.

  \subsection{Isochrone interpolation - the grid formalism}

The first step in our method is to generate a fine grid of isochrones, with
age and metallicity steps significantly smaller than the observational
error. This becomes the basis for the simulated stellar population.  One
then generates a population distribution in age and metallicity, fixed
discretely to these grid points.  Each grid point corresponds to one
isochrone, so that after the population has been created, populating a
complex system becomes a matter of cycling through each grid point,
generating a list of stellar masses using the given Initial Mass Function
(IMF), the mass ranges of that particular isochrone and the relevant
observability constraints.

\begin{figure}
\vspace*{7.5cm}
\includegraphics{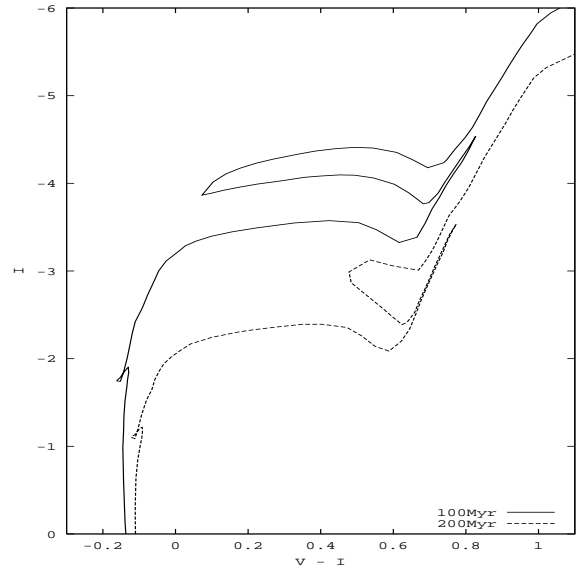}
\caption{An example of isochrones showing sharp `critical turning points'.
Isochrones are for 100Myr and 200Myr, with a metallicity of one-fifth solar.
Isochrones from Girardi et al. (2000)}
\label{turningpoints}
\end{figure}

We use the technique of automatic critical point analysis in order to
interpolate accurately between adjacent isochrones (Fig.
\ref{turningpoints}).  This method allows us to identify the parts of the
isochrones which are most important, and pair them up so that these features
are carefully preserved.

Girardi et al. (2000) identify the more rapidly changing positions in their
isochrone set, but in general this information is not available except by
inspection.  In this work we identify these points automatically by
considering sharp changes in the tangent gradient to the isochrone curve. We
consider here the $V$, $V-I$ plane, though future studies will also include
other information available in the isochrone data, such as plots of $log(L)$
vs. $log(T_{eff})$, in which these loci are more distinct.

When the critical points have been identified, we must pair them up and then
interpolate between them.  At this stage it is necessary to calculate the
extent to which we must interpolate between the available isochrones.  This
is mainly a trade off between the computational time required to interpolate
the isochrones and the requirements for the accuracy of the CMD.

Clearly, if we choose a fairly coarse isochrone spacing in age and
metallicity, then this limits our sampling accuracy.  Correspondingly it is
wasteful to store isochrones to extremely high precision in
metallicity and Age.  A sensible balanced medium of these two cases is
required.

Once the isochrone set has been fully interpolated then we can produce
artificial stellar populations using it.  In our code, an input central
metallicity and age is specified for each population, together with Gaussian
spreads in each quantity, and the desired photometric errors, population
size, completeness limit and IMF.  Stars are generated according to the
input distribution, and then their details are calculated by clipping them
to the nearest isochrone model and reading off the desired observational
properties, interpolating in mass as required.

  \subsection{Introducing errors}
  \label{photometry}

Often we require a simulation of a stellar population as it would be seen
from a significant distance, for example the halo of a nearby galaxy.  In
this case it is important to add in completeness effects so that the number
of stars tails off to fainter absolute magnitude rather than increasing with
the background correspondingly increasing.

An indication of the form of a completeness function can be derived
analytically for the ideal case of isolated sources on a smooth background.
For ease of calculation, and due to the low importance of the exact form of
the completeness function to our methods, we decided to model this effect as
a cutoff varying linearly between $\pm 1 mag$ above and below the
completeness magnitude.  This value is either specified manually or
determined experimentally by examining the luminosity function of the
low-magnitude stars in the data set.  The exact value of this limiting
luminosity is not of great importance to the subsequent analysis.

We also introduce random photometric errors into the dataset, using a
user-defined scale value set at run-time and a Gaussian smoothing kernel.


\section{Fitting metallicity distributions to CMD data}

The process of obtaining star-formation histories from CMD data has
produced a great deal of literature in recent years.  Amongst them are
reviews and analyses by Aparicio et al. (1996), Tolstoy \& Saha (1996),
Dolphin (1997) and Hernandez, Valls-Gabaud \& Gilmore (1999).

The main important differences between all of these methods and that
outlined here is the depth of the photometry involved.  All four of these
examples consider photometry down to the Main-Sequence Turn-Off (MSTO)
magnitude.  Our method was designed to be applicable mainly to the RGB
stars, and is therefore optimised for fitting old, evolved populations where
the age-metallicity degeneracy still causes considerable problems.  We
therefore cannot consider our input isochrones to represent a set of unique
{\it eigenpopulations}, but rather a spanning, but oversampled set.

In addition, we have focused on the optimisation procedure. Most methods
have adopted relatively simple methods for discovering the best-fitting
model coefficients.  These do not address the risk of completely missing a
globally optimum solution in complex CMDs, with the exception of the method
of Hernandez et al. (1999).  Harris \& Zaritsky (2001) consider a
gradient-walker method, as outlined below, with safeguards against
converging towards a local minimum.  Our methods differ from theirs in three
main areas.  Firstly, we consider several different methods both for
generating probability matrices and for optimising coefficients, testing each in
turn.  Secondly, we analyse the degree to which each method avoids falling
into only local minima.  Thirdly, we start our optimisation procedure with a
first guess coefficient vector rather than a random initial guess, which
goes some way to alleviating this problem.

In the following sections we discuss several methods for generating
probability matrices and optimising model-fitting coefficients, and explain
the advantages and disadvantages of each.

Finally, we introduce a new method for calculating the isochrone fitting
coefficients for each star that leads to an accurate output metallicity
determination which is robust to a real (model) dispersion.

  \subsection{Methods}

In order to analyse any CMD data it is necessary to compare the loci of
stars detected in the given CMD to the positions of theoretical isochrones.
This is the principle of `isochrone fitting'.

There is an inbuilt complexity with isochrone fitting, namely the
age-metallicity degeneracy.  The effects of increasing metal abundance on
stellar isochrones are remarkably similar to those of increasing age.  In
fact, with the absence of any other diagnostic methods, it becomes
impossible to separate the two effects in some parts of the CMD, hence the
degeneracy.  For a population of age 2Gyr and metallicity of one half solar,
doubling the metallicity has a very similar effect on the locus of the RGB
as an increase in age to 7Gyr. The most important place on the isochrone
where this is not true is at the MSTO.

At the MSTO, the effects of age and metallicity are different, and
vary depending on the wavelength at which they are examined. 

However, the MSTO is significantly below the tip of the RGB, by
approximately 6 magnitudes in $I$ for an intermediate age population, and
significantly more for older populations.  Clearly extremely deep images are
required to achieve this kind of photometry in external galaxies, and this
simply is not possible for galaxies far outside the Local Group.  For the
majority of targets it is therefore impossible to achieve sufficiently deep
photometry to resolve the MSTO, and the degeneracy remains.

  \subsection{The old-age restriction.}

This study deals primarily with the analysis of Population II stars in
galaxy haloes.  To analyse intermediate age stars we need to use different
techniques.  The use of the AGB can help to place restrictions on the ages
of intermediate-age stars, but this is of no help for stellar populations
older than 7 or 8 Gyr.  At older ages, the AGB is much fainter and the main
diagnostic tool we have is the RGB.  Fortunately, at old age, the effects of
age on the RGB are minimal, and the dominating parameter is metallicity.

\begin{figure}
\vspace*{7cm}
\includegraphics{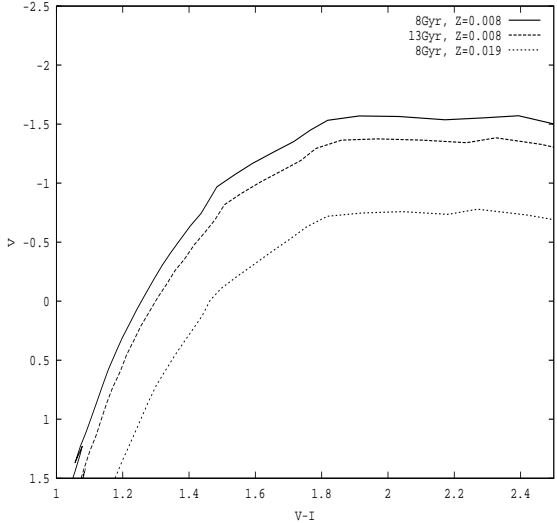}
\caption{The difference between RGB isochrones of different ages is small at
old ages.  Metallicity, however, remains an important 
discriminator.  Here we have plotted one isochrone for an 8Gyr old
population with half solar metallicity (solid line).  The difference caused
by increasing the age by 8Gyr to 13Gyr (dashed line) is not particularly
large, but that caused by increasing the metallicity to solar abundance
(dotted line) is far greater.}
\label{oldage}
\end{figure}

Figure \ref{oldage} demonstrates the difference between age and metallicity
variation at high ages.

Notice how the age increase of more than 50 percent has very little effect
on the colours and magnitudes of stars, whereas the metallicity increase
from half-solar to solar abundances has an enormous effect.  For populations
of 10Gyr or older, an age variation of a few Gyr is important at
approximately the same level as a metallicity variation of 0.07 dex at solar
abundance.  This causes a degeneracy in the metallicity distribution at
approximately this level for a population with a 2Gyr age spread.  For more
coeval populations with narrow age spreads, the metallicity
smearing effect is even lower.

Though this does not break the degeneracy per se, it certainly alleviates the
problems it causes.  Old RGB stars can be analysed with an assumed age, and
a fairly accurate metallicity distribution can be obtained.  Note that this
method only applies to the RGB.  By the time one reaches the horizontal
branch then second parameter effects render this method useless.

Hence, the metallicity distributions that we recover using our methods are
only approximations to the true form, assuming one single age.  For old
populations, our results will be a good approximation to the metallicity
distribution of the stellar population as a whole, regardless of age spread.
For populations with younger components, the true metallicity values
recovered should not be taken quite so literally.  Instead, what we can
recover is the overall {\it form} of the metallicity distribution, allowing
us to identify the existence and relative sizes of any low- and
high-metallicity components, though not their precise metallicity values.

This method for obtaining metallicity spreads is naturally dependent on
obtaining photometry sufficiently deep to avoid biassing against the
higher-metallicity components.  Clearly from Fig. \ref{oldage}, photometry
only complete to -1.5 in $V$ would introduce a strong bias against
higher-metallicity populations.

In addition there will be a slight bias due to the fact that populations of
different metallicities evolve at slightly different speeds, so therefore we
expect the size of the RGB population as a fraction of the entire population
to vary with metallicity.  Thus, if we have shallow photometry covering just
the RGB we expect to make errors of the order of 10\% in the relative
weighting of metal-poor populations over metal-rich populations.  This is a
minor effect compared with other sources of numerical error.

\section{Bayes' Theorem \& Bayesian Inference}

Bayes' theorem allows one to calculate the probabilities of certain models
having created an observed data set.  This probability can be maximised in
order to discover the most likely model, or formation scenario for those
data.  Bayesian Inference is vital in determining the most probable
distribution of metallicities in any stellar population.  A more rigorous
review of this process is found in Tolstoy \& Saha (1996).

Consider a set of observed data, $A$, together with a set of models, $B$,
from which we presume all of the observed data could have been derived. The
problem we face is to find the most probable model, $B_{best}$, in a rigorous
way. Essentially, which of these models, $B_n$, has the highest probability
of producing the observed data, $A$?

We introduce the nomenclature $A_n$ being the $n^{th}$ element of the data
set $A$.  $B$ is defined as the set of all possible models, with each model
$B_n$, comprising a linear combination of (non-orthogonal) basis elements.

The probability of an observed data set, $A$, being observed and $B_n$ being
the model which created it is simply given by;

\begin{equation}
P(A,B_n) = P(A|B_n) \times P(B_n)
\end{equation}

From this, it is trivial to derive Bayes' theorem, which states;

\begin{equation}
P(B_n|A) = \frac{P(A|B_n) \times P(B_n)}{P(A)}
\label{bayes}
\end{equation}

In other words, the probability of a certain model, $B_n$ being the true
model, given the observed data $A$, is related to the probability of
obtaining those data $A$ from model $B_n$ and the prior probability of model
$B_n$, together with a normalisation factor.

This normalisation factor, $P(A)$, is straightforward to calculate.  In the
special case where the models $B_n$ are exclusive and exhaustive, it
reduces to the following;

\begin{equation}
P(A) = \sum_{n} P(A|B_n) \times P(B_n) = C
\end{equation}

where C is a constant.  Thus we can rewrite equation (\ref{bayes});

\begin{equation}
P(B_n|A) = \frac{P(A|B_n) \times P(B_n)}{C}
\end{equation}

$P(B_n)$ is known as the prior distribution, and allows us to bias the
probabilities if we know that one particular model is $a~priori$ more likely
than another.  Essentially, if we calculate that a dataset is equally likely
to have come from two separate distributions but that one of the
distributions occurs more often in nature, then without any further
contradictory evidence, it is safe to assume that this is the more probable
distribution to have created the observed data.

In the astronomical context, we have very little prior knowledge of the
metallicity distribution functions involved.  Initially, it is safe to
assume that all models are equally likely, though later we will introduce the
concept of penalty functions, attempting to impose a certain form to the
distribution function such as smoothness or a Gaussian profile.

  \subsection{Model fitting}

We have obtained an expression telling us how we discover the most probable
model given the data available.  Assuming the $P(B_n)$ are all the same, as
above, then we have;

\begin{equation}
P(B_n|A) \propto P(A|B_n)
\end{equation}

So the problem of finding the most probable input distribution reduces
initially to one of finding the distribution which maximises the likelihood
of creating the given data.  Of course in reality we cannot possibly
consider the totality of all possible models, as this is prohibitively
large, but we can only consider a subset.  It is necessary, then, to
consider a sensible subset that is guaranteed to include a good
approximation to the globally most likely model, and as little else as
possible.  Of course, $a~priori$ we have no idea what the most likely model
is, so we consider a sensible subset of all models depending on the problem.

In this work, we wish to fit a metallicity distribution to an observed
stellar population with an observational constraint that this population is
`old'.  We have already generated a large library of isochrones with which
this hypothesis can be tested.  The problem reduces to the following;

{\em What linear combination of isochrones of varying metallicity and age
best reproduces the distribution of stars in an input dataset?}

In fact, we have considered only a subset of this question.  We know that
the Age-Metallicity-Distance degeneracy prohibits an absolute solution to
this problem without unobtainably deep data.  Instead we adopt in turn a set
of single ages, initially 12Gyr, and then fit a metallicity distribution
assuming each age in turn.  The degree to which the stars cannot be fit
using this assumption tells us a little about the degree to which the coeval
assumption is incorrect.  We vary the assumed age, and study the variations
in the derived metallicity profile.

That is to say, we need to find the optimum set of coefficients, $\alpha_n$,
to maximise the product;

\begin{equation}
L = \prod_{i} \Bigg( \sum_{j} \alpha_j P_{ij} \Bigg)
\label{likelihood}
\end{equation}

where $P_{ij}$ is the probability that star $i$ was obtained from
isochrone $j$.  The calculation of this matrix is discussed in section
\ref{isofit}, and the process of optimising the coefficients is discussed
in section \ref{optimisation}.

\section{Isochrone fitting}
\label{isofit}

The first implementation problem we encounter is that of calculating the probability
matrix, $P_{ij}$, which lists the probability of star $i$ having come from
isochrone $j$.  Clearly we need an accurate measure of this quantity, which
will allow us then to optimise the weightings for each isochrone such that
the overall probability of the observed distribution coming from the given
combination of isochrones is maximised.

  \subsection{Determining the probability matrix, $P_{ij}$}

The fundamental part of the optimisation process is determining an accurate
probability matrix, $P_{ij}$.  To do this, we require a method of
calculating whether or not one star came from a particular isochrone, based
only on a discrete parameterisation of the isochrone and the observed
position of the star, together with any known observational errors.  Clearly
for a theoretical population, the only errors are those caused by numerical
problems, rounding off problems etc.  However, for a real population,
we need to account for the fact that real stars do not behave optimally, and
also that real detectors are not perfect.  Therefore, there is a finite
measurement error in both $V$ and $I$.

That is, just because a star does not {\em appear} to lie on a particular
isochrone, we cannot be sure that its displacement is not entirely caused by
observational errors.  Also, we must account for the fact that
our isochrone models are not exact representations of observed RGBs, as
discussed earlier.

  \subsection{Gaussian magnitude errors}
  \label{gaussian}

It is useful to consider a set of N points arranged in a CMD at positions
$\gamma_n = (\gamma_n^x,\gamma_n^y)$.  Associated with each of these points
is an error in each of $x$ and $y$, or alternatively the magnitude and colour
parameters.  We could just as easily work with the two individual magnitude
parameters, such as $V$ and $I$, as a magnitude-magnitude diagram (MMD)
contains exactly the same information as a CMD, just arranged in a different
manner.

Regardless of the parameterisation, we designate each point as $\gamma_n$,
and the associated errors on this measurement as $\delta_n =
(\delta_n^x,\delta_n^y)$.  Thus we associate a probability density
distribution, $\rho(x,y)$, with every point in space. If we assume that the
probability density distribution is Gaussian in form then we can assign a
bivariate Gaussian probability function to $\rho$ thus;

\begin{equation}
\rho(x,y) = \frac{1}{2 \pi \delta_n^x \delta_n^y} exp \bigg[ - \frac{1}{2} \bigg\{
\frac{(x-\gamma_n^x)^2}{{\delta_n^x}^2} + \frac{(y-\gamma_n^y)^2}{{\delta_n^y}^2}
\bigg\}\bigg]
\label{probdef}
\end{equation}

This is normalised so that the integral over all space

\begin{equation}
\int \int_{-\infty}^{\infty} \rho(x,y) dx dy \ = \ 1
\end{equation}

If we sum over all of the points in our distribution, we obtain the
total probability density at a point at position $\mathbf{r} = (x,y)$.  We
designate this as $p(\mathbf{r})$;

\begin{equation}
p(\mathbf{r}) = \sum_{i} \rho_i(x,y)
\label{probdef2}
\end{equation}

where $\rho_i(x,y)$ is the contribution to $\rho$ from the $i^{th}$ point in
the distribution.

We now apply this to isochrones, introducing the concept of an isochrone
parametrised as a number of discrete points, rather than a continuous
distribution.  The more points we have, the closer this approximation
becomes.

This determines the probability that any observed star came from a specified
isochrone, $j$ as $P_{ij}$.  That is defined from equation (\ref{probdef2});

\begin{equation}
P_{ij}(\mathbf{r}) = \sum_{i} \rho_i^j(x,y)
\label{probiso}
\end{equation}

where $\rho_i^j$ is the probability density function arising from the
$i^{th}$ star in isochrone $j$.

The probability that some distribution of stars arose from a certain
isochrone is;

\begin{equation}
\xi_n = \prod_{i} P_{ij}
\label{probdist}
\end{equation}

and therefore for some linear sum of isochrones, with coefficients
$\alpha_j$, we derive the likelihood $L$ that some distribution of
stars, labelled as $i$, came from the distribution of isochrones labelled
$j$.  This is the result stated in equation (\ref{likelihood}) above.

Of course, this likelihood is only an expression of the {\bf unnormalised}
probability distribution, and we can not obtain any absolute estimate on the
quality of our model fits simply from this value.  However, it gives us a
relative measure whereby we can compare any two distributions and assess
which is a `better fit' to the data.

In addition, we maximise the natural logarithm of this quantity,
that is;

\begin{equation}
ln L = \sum_{i} ln \bigg( \sum_{j} \alpha_j P_{ij} \bigg)
\label{loglikelihood}
\end{equation}

This is bijectively related to the likelihood in the range under
consideration, with no change of ordering, that is;

\begin{equation}
\forall \{a,b\} \in \Re^{+},   a < b \Rightarrow ln(a) < ln(b)
\end{equation}

So that the operation of $ln$ is order-preserving.  This means that
maximising this logarithm is equivalent to maximising the likelihood itself.

It is important that the isochrone is sufficiently finely interpolated so
that the spacing between points is of the same order of the Gaussian errors,
or less.  This ensures that the final answer is not a strong function of the
actual parameterisation of the isochrones, and is an accurate
representation.  In practice, it is sensible to allow interpolation of the
input isochrone to arbitrary fractions of the Gaussian errors in the
required colours.

The summation need be carried out only over a small subset of the total
number of isochrone points in practice.  This is because the value of a
Gaussian probability distribution rapidly drops after a few error radii, and
by the time either one of the displacements reaches 7 or 8 $\sigma$ then the
probability is, to all practical purposes, zero. Therefore a binary search
was used to find the nearest isochrone point to the observed CMD point, and
a restricted number of points in the vicinity of this one point were
included in the calculation.

In this way, all the points which were capable of contributing a
non-negligible fraction to the likelihood were calculated, with a minimum of
extra stars outside the tolerance radius.  This reduced the order of this
operation from $O\big(n\big)$, where $n$ was the isochrone length, to a
constant independent of $n$, depending only on the level of isochrone
interpolation.  The binary search was extremely fast, and though it had a
greater (logarithmic) time complexity, its contribution to the computation
time was always negligible.

This bivariate Gaussian method has several advantages over other similar
methods:

\begin{enumerate}
\item The precision of this method is defined by the accuracy of
the isochrone interpolation.  The more accurate one requires the
probabilities, the more finely one needs to interpolate the isochrone.
\item This method has a strong statistical grounding, building
only on known properties of the error distributions involved, both those due
to observational techniques, and those intrinsic to the distribution under
study.
\item The bivariate Gaussian probability distribution is robust, if
slow, to calculate.
\end{enumerate}

However, it does introduce some unwanted problems:

\begin{enumerate}
\item Discrete parameterisation of the isochrones introduces an extra spread
in probabilities due to the fact that some points will, purely by chance,
fit less suitable isochrones.
\item Calculating a large probability matrix is rather slow, taking several
minutes for a heavily populated CMD and a finely interpolated isochrone set.
\item Further interpolation of the isochrones is necessary, though
theoretically this gives us no more information than it originally contains,
thus this method is inefficient.
\end{enumerate}

\begin{figure}
\vspace*{3cm}
\includegraphics{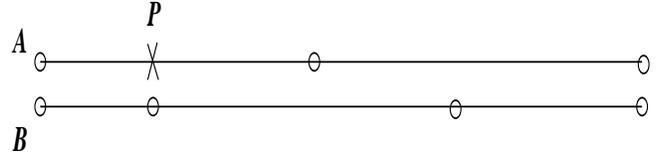}
\caption{The problem of assigning stars to the wrong isochrone segment.}
\label{misassignment}
\end{figure}

An illustration of the first of these problems is shown in figure
\ref{misassignment}.  The point labelled $P$ actually comes from isochrone
$A$, but the Gaussian probability method described above would assign a
larger probability of this point belonging to isochrone $B$, simply because
of the discrete parameterisation.  Of course, this problem is alleviated
somewhat by a sufficiently fine isochrone interpolation.

\vspace*{0.5cm}

\vspace*{0.5cm}

We also investigated other methods of generating the probability matrix.

  \subsection{The Perpendicular Distance Method}
  \label{perpendicular}

A second method for generating the probability density matrix, $P_{ij}$, is
the perpendicular distance method.  This is potentially more accurate,
though slightly more difficult to implement.  The method was formulated in
order to avoid the problems described above due to the discrete nature of
the isochrones and the possibility of this causing mis-classification
problems. A method was desired which was more information-efficient than the
Gaussian method described above, and also faster to calculate.

The obvious solution to the problem was to abandon the Gaussian distribution
of points, and instead calculate how far any particular star is from each
isochrone in a perpendicular sense.  In other words, calculate the closest
distance between any star and each isochrone, and assign a probability based
on just these distances, rather than a sum over all isochrone points.

Actually calculating the nearest distance is simple. Some geometry gives the
following relation;

\begin{figure}
\vspace*{6.5cm}
\includegraphics{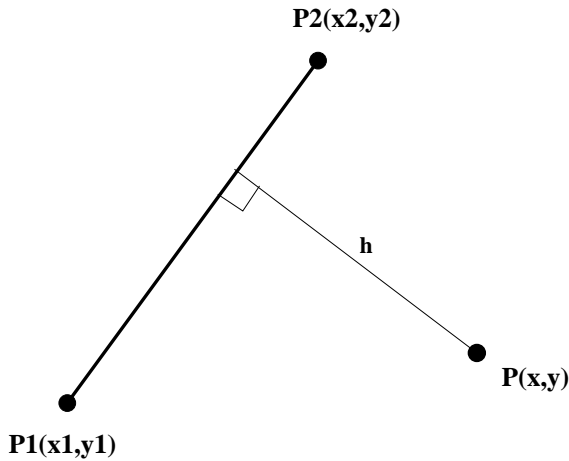}
\caption{Geometry for the perpendicular distance value.}
\label{geometry}
\end{figure}

\begin{equation}
h = \frac{2}{R} \Big((x-x_1)(y_2-y_1) - (y-y_1)(x_2-x_1)\Big)
\label{perpendeq}
\end{equation}
  
where 

\begin{equation}
R = \sqrt{(x_2-x_1)^2 + (y_2-y_1)^2}
\end{equation}

The only real problem is how to determine the nearest two points, designated
$P_1$ and $P_2$ in figure (\ref{geometry}).  Initially it seemed easiest to
use the nearest isochrone point to the star under consideration, plus the
nearest of its two immediate neighbours in the isochrone.  In practice this
led to many problems, so we decided to parse through the entire isochrone,
calculating the perpendicular distance for each pair of points, and then
finding the minimum value, after applying mass-weightings and
cutoff-weightings (see section \ref{weighting}).

Now the perpendicular distance has been calculated, it is a simple matter of
calculating the Gaussian probability density of this value. This is given
by;

\begin{equation}
P_{ij} = k e^{- \frac{h^2}{2 \sigma^2}}
\end{equation}

Where $\sigma$ is a combined error radius, and $k$ is a constant defined to
give answers in a reasonable range.  Clearly, we are uninterested in the
exact value of $k$ as only relative probabilities are considered.  Provided
the same value of $k$ is chosen for all calculations then the relative
probability ratios will be the same regardless of the exact value chosen.

It is, of course, possible to split up the perpendicular vector into two
components parallel to the colour and magnitude axes.  This means that
different errors in the two filters can be considered without any added
difficulty simply by using a bivariate Gaussian distribution to assign the
probability instead of the above method.

  \subsection{Mass weighting and cutoff weighting}
  \label{weighting}

Part of Bayes' theorem states that all models are equally likely in the
absence of any prior information to the contrary.  Clearly, introducing more
information known to be intrinsic to the problem under study can only
improve our chances of retrieving a sensible and accurate answer, provided
that such information is indeed sensible and accurate.

In any stellar population, stars are produced with a certain IMF.  For the
RGB, most stars have the same mass to within a very small fraction.
However, this paper describes a general procedure which can be applied to
stars of any mass or luminosity, in principle.  We therefore adopt the
Kroupa, Tout \& Gilmore (1993) mass function, which takes the following
form;

\begin{center}
\begin{equation}
N(m) \ \propto \ m^{-\alpha}
\end{equation}
\end{center}

with the following 3-element power law slope;

\begin{equation}
\left.
\begin{array}{lll}
\alpha = 2.7 & when & \textrm{$m > 1.0 M_{\odot}$}\\
\alpha = 2.2 && \textrm{$0.5 \le m \le 1.0 M_{\odot}$}\\
0.7 \le \alpha \le 1.85 && \textrm{$0.08 < m < 0.5 M_{\odot}$}\\
\end{array} \right.
\label{ktg_imf}
\end{equation}

Taking this information, we can apply it to the calculation of the
probability matrix, $P_{ij}$.  If we have a star which is equidistant from
two separate isochrones, then we can calculate the mass that this star would
have at the nearest point in each isochrone, and bias the probability by
this.  Clearly the chances are that it came from the isochrone for which
this interpolated mass is smallest.

Therefore we can bias the probability of each star coming from each
isochrone by the IMF factor calculated as above in equation (\ref{ktg_imf}).
The IMF has also to be taken into account for the normalisation in the
Gaussian approach, section \ref{gaussian}.  However, in the Perpendicular
Distance method of section \ref{perpendicular} we need not do this.

We may also bias the probability by the completeness function for the point
at which the perpendicular displacement touches the isochrone, or the
nearest isochrone point to the star under question.  Of course this is not
possible with the Gaussian method, as we don't calculate the luminosity of
the perpendicular intersection points.

\begin{figure}
\vspace*{5.5cm}
\includegraphics{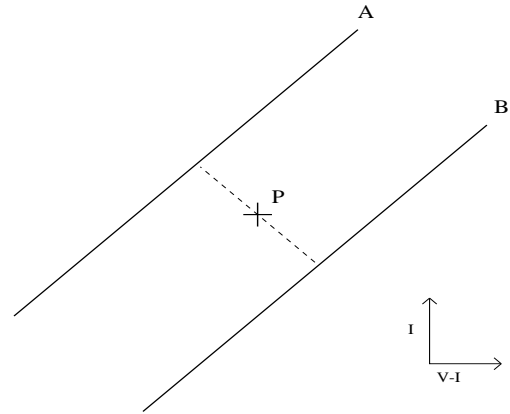}
\caption{An illustration of how completeness effects can bias probabilities
between otherwise equidistant isochrones.  Here, the star is more likely to
have come from the isochrone whose intersection point is at a brighter
magnitude.}
\label{cutofffig}
\end{figure}

In figure \ref{cutofffig}, the star $P$ is equidistant from two isochrones,
but the perpendicular displacement towards isochrone $A$ touches at a
brighter apparent magnitude, so we can say that this star is more likely to
have come from isochrone $A$ than isochrone $B$, all other effects being
identical. This works only for stars around the completeness limit, of
course.  It is contrary to the bias introduced by the IMF.

  \subsection{Comparing the methods}

The Perpendicular Distance method has several advantages;

\begin{enumerate}
\item Probabilities are independent of the exact position of the
isochrone points, provided that the isochrone is sufficiently accurately
interpolated in the first place.
\item No further interpolation of the isochrone is required.  Therefore this
method is optimal in terms of information efficiency.
\item This method is much faster than the Bivariate Gaussian sum described
in section \ref{gaussian} above.
\item Because of the way this method treats stars lying very close to
isochrones, where the perpendicular displacement is less than the average
isochrone point separation, the probability gradient is much steeper than
for the Gaussian method.  This makes it easier to reject particularly
unlikely models, especially in datasets with small photometric errors.
\end{enumerate}

However, it does also have some drawbacks;

\begin{enumerate}
\item This method is not as easy to enhance accurately with mass-weightings.
See section \ref{weighting} for details.
\item Linear interpolation has a larger effect on the errors here than it
did in the Gaussian method because we are dealing with just one measurement,
rather than the sum of several, and therefore a substantial fractional error
could prove significant to the overall probability.
\item Most importantly, this method is susceptible to end-point errors.
\end{enumerate}

It is worthwhile considering the last two of these problems in turn.

    \subsubsection{Linear interpolation problems}

In this method, the effects of linear interpolation between points are more
pronounced.  With the Gaussian error box method of section \ref{gaussian},
we knew that at least some of the points we considered were at the correct
place in the isochrone.  However, in this method, we use the actual
calculated isochrone points only for {\it reference}.  The perpendicular
distance derived is one taken from linear interpolation between these two
points. There is no component taken directly from the isochrone
parameterisation, so our probability is entirely reliant on the accuracy of
the interpolation.

It is easy to develop contrived examples where linear interpolation produces
enormous errors.  Consider, for example, the case of a sharp pointed corner
in the isochrone where linear interpolation completely cuts off the full
extent of the corner, and therefore introduces a large error on any points
found near this region.

Of course, with artificial datasets this problem will have no effect
whatsoever.  The artificial sets are generated using the same linearly
interpolated isochrones.  It is slightly more worrying therefore that the
extent to which this problem affects the results cannot easily be measured.
However, several facts suggest that this problem should not be too severe:

\begin{enumerate}
\item Interpolation problems mostly occur around stages of rapid stellar
evolution, which are rare.  In analysing mainly the RGB we avoid most such
loci.
\item These stages of rapid evolution are much more carefully analysed in
the simulations which produce the isochrones, so the isochrone points are
closely spaced at these places.  This means that any errors are likely to be
kept to a minimum.
\item These are not systematic errors, and are distributed in an
approximately random fashion on either side of the desired isochrone.  The
optimisation methods discussed below in section \ref{optimisation} smooth out
the occasional anomaly.
\end{enumerate}

Other linear interpolation problems were also considered, such as the
problem where the nearest isochrone point to a location in CMD space is at a
sharp turning-point in the isochrone.  Fortunately testing for this is
simple, so methods were developed to correct for the problem.

  \subsection{Variable photometric errors}
  \label{perrors}

As with the treatment of errors in section \ref{photometry}, it is necessary
here to consider variation of photometric errors over the full spread of
colour-magnitude space.  Most importantly, the expected value of the
photometric error is much larger near the completeness limit as $\sqrt{n}$
and confusion effects begin to have a more considerable impact.

For the Bivariate Gaussian method, the application is rather
straightforward.  For each point on the isochrone, an associated error in
each of the two colours is calculated based on the global error value and
the distance from the cutoff in that band.

For the perpendicular distance method, the solution is rather more
complicated, and requires calculation of the components of the perpendicular
displacement vector in each of the orthogonal directions representing the
magnitude and colour. Computationally this is rather slow, but need only be
determined once for the assignment of initial probabilities.


\section{Coefficient optimisation}
\label{optimisation}

Once the probability matrix, $P_{ij}$ has been generated, the next step is
to optimise the coefficients $\alpha_j$, to maximise the logarithm of the
likelihood, as described in equation \ref{loglikelihood}.

This is a constrained maximisation in a space with dimensionality determined
by the number of isochrones available.  The constraints are;

\begin{enumerate}
\item The coefficients are constrained to $0 \le \alpha_j \le 1 \ \forall j$
\item The sum of the coefficients is constrained to unity, $\sum_{j}
\alpha_j = 1$
\end{enumerate}

  \subsection{Background distributions}
  \label{background}

As already discussed, one encounters problems with anomalous stars in the
above method.  For stars that lie sufficiently far away from all isochrones
to be considered unrelated, numerical problems occur.  That is to say, for a
star $i$ such that $P_{ij} = 0 \ \forall j$, one obtains terms from the
logarithmic likelihood equation, (\ref{loglikelihood}), of logarithms of
zero. One way of dealing with this problem is simply to test for zeros and
before the logarithm is taken return a very large negative value.

However, a more physically reasonable work around is to introduce a
background distribution which attempts to fit all of the anomalous stars and
remove them from the problem.  This stops the related problem where a star
is far from all isochrones except one, and where the weighting for this one
isochrone is then increased just simply to enable it to fit this one wayward
star.

A sensible assumption in any real dataset is that a certain proportion of
the stars are either foreground stars, background objects such as distant
galaxies, or simply spurious.  Clearly one does not wish to fit these extra
objects and trying to do so would confuse the isochrone weighting algorithms
thereby producing incorrect results.  A more sensible solution is to just
estimate this background fraction, and then introduce a new model, in
addition to the isochrones, which essentially represents an isotropic
probability density distribution of background objects over all of
colour-magnitude space.  All stars are guaranteed to fit this distribution,
so if we call it model zero, we obtain $P_{i0} = 1 \ \forall i$.

If we insert this into the equation for the logarithmic likelihood, we then
obtain the following analytic form;

\begin{equation}
log L = \sum_{i} log \bigg( \sigma + \sum_{j} \alpha_j P_{ij} \bigg)
\label{loglikelihoodbg}
\end{equation}

Here, $\sigma$ is defined as $\alpha_0 P_{i0}$, which is equal to $\alpha_0$
independently of $i$.  Therefore, $\sigma$ is equal to the weighting of the
background proportion, and hence affects the degree to which anomalous stars
are fitted.  This therefore requires slightly different constraints on the
normalisation of $\alpha$ if we require the total model weighting to sum to
unity.  Now we require

\begin{equation}
\sum_{j} \alpha_j = 1 - \sigma
\label{norm}
\end{equation}

It is worth noting that we fix the value of $\sigma$ here for one very good
reason.  Clearly, we could allow the maximisation algorithm to vary the
weighting for the background distribution in order to fit it as best it
could.  However, then it would have freedom to increase the weighting to 1,
and reduce all other weightings to zero, because the background distribution
is guaranteed to fit all stars, by definition.

One could equally well introduce a functional form for the background
distribution where it varied depending on the magnitude or colour of the
stars it was fitting.  In that case it would be necessary to precalculate
the values of $P_{i0}$, which would now no longer all be equal to unity, but
would vary with $i$. This method would be useful for fitting background
stars in fields which are known to be contaminated with, for example, faint
blue background galaxies, or perhaps a foreground star cluster, and where
reliable offset data are available.  Another interesting possibility would
be to simulate the background distribution for fields close to the Galactic
disk, using a simple galactic model.

  \subsection{Smoothness constraints}
  \label{smoothness}

As suggested earlier in section (\ref{isofit}), it is possible to introduce
any prior information about a particular field, or the functions under
consideration, in order to restrict the range of the fit.  One further
consideration is the smoothness of the metallicity function.  This can be a
desirable property.

For example, we might not expect spiky metallicity distributions, but rather
a much smoother shape, or vice versa.  Introducing these constraints into the
likelihood function allows us to optimise the fitting procedure in either
situation.

The implementation is numerical rather than analytic.  We introduced code
into the likelihood calculation function that penalises large steps in
$\alpha$ and also reduces the likelihood depending on the number of
individual maxima in the metallicity distribution, hence selecting against
particularly rough distributions.  In complex stellar populations, one
expects a certain degree of roughness e.g. caused by late accretion events.
Smoothing levels must be tested to ensure that the metallicity distributions
returned are not losing real information.

A second method is to estimate the expected uncertainty in the metallicity
values obtained, and to convolve a Gaussian of this width with the final
metallicity distribution.  Our second paper will deal in more detail with
the estimation of intrinsic errors due to distance and age uncertainties.

\section{Maximisation algorithms in multi-dimensional parameter space}

We have now developed a method for calculating the degree to which any
distribution is fitted by a given linear combination of isochrones,
parameterised by coefficients $\alpha_i$.  All that remains is to optimise
these coefficients using some algorithm to be determined.

The sections are labelled in order of increasing complication, with the
latter methods being both more difficult to fine-tune, but also more
successful when working correctly.

As a first guess, it is accurate enough simply to analyse the probability
matrix, and build up a normalised set of coefficients based on how many
stars each isochrone fits best.  One loops through each star, and calculates
which isochrone has the highest probability of producing this star, that is
to say find the value $j$ for a star $i$ which maximises $P_{ij}$.  A tally
of how many stars are best fit by each isochrone is made.  Once all stars
have been considered, we normalise the tally vector and set this as the
initial coefficient vector.

Clearly this is a good first guess, but there are problems with this method.
Most obviously, in places where the isochrones are crowded, it is quite
possible that a star will be best fit by an isochrone which is not the
correct isochrone, and is not even adjacent in metallicity.  This problem is
enormously reduced by using the perpendicular distance method as discussed
above.

  \subsection{The Gradient Walker Algorithm}

To visualise the problem, it will prove useful to imagine the simplified
case of maximising a function of two variables, and then simply to extend
this mathematically to a larger number.  One can readily imagine a height
field where the height $h(x,y)$ is a function of the two orthogonal
cartesian displacements $x$ and $y$.  The problem is to find the
maximum value of the function $h$ given the constraints that $x \ge 0$ and
$y \ge 0$ and also that $x+y=1$.  If we forget about the third of these
constraints for a moment, we can imagine the problem as maximising the
function $h$ over a region of the $x,y$ plane.

In many dimensions, we can generalise this problem to that of maximising a
function of $n$ variables with $n+1$ constraints.  Namely, that each of the
separate variables must be positive, and that the coefficient vector,
$\mathbf{\alpha}$ is constrained to lie within the hypersurface defined by
$\sum_{j} \alpha_j = constant$.

The most obvious method of finding the maximum of this function is a
so-called {\em gradient walker}.  That is to say, one takes a series of
steps where at each point the gradient of the function is evaluated, and
then some scalar multiple of this vector is added to the coefficient vector.
The coefficient vector is then renormalised and the new likelihood is
calculated.  This method is repeated until a maximum is found.

    \subsubsection{Calculating the gradient}

One can in principle calculate the gradient of the likelihood function
easily.  The following is the appropriate formula to use;

\begin{equation}
\frac{\partial log(L)}{\partial \alpha_k} =
\frac{log\big(L(\alpha + \epsilon_k)\big) -
log\big(L(\alpha)\big)}{\epsilon}
\label{gradient}
\end{equation}

where $\epsilon_k$ is some small incremental vector equal to the following;
  
\begin{equation}
(\epsilon_k)_j = \left\{ \begin{array}{ll}
\epsilon & \textrm{if $j=k$}\\
0 & \textrm{if $j \neq k$}
\end{array} \right.
\end{equation}

This calculation has to be performed for every dimension, essentially once
per isochrone.  When there are several hundred, or even thousands of
isochrones to fit then this method becomes costly.  However, the expectation
is that this calculation need only be performed a few times before the
maximum is reached.

    \subsubsection{Variable step length}
    \label{vsl}
  
Assuming the gradient vector can be calculated, it is then normalised to unit
length and multiplied by a scale factor depending on how much we wish
to alter the coefficients.  This value is very much open to fine-tuning,
and it is principally a matter of experimentation to discover the optimum in
any specific application.

Of course, the gradient step can, and should be altered depending on how
close one estimates one is to the maximum of the likelihood function.  This
can be approximated by the following method, using the data from the last
step taken;

\begin{equation}
Improvement \ = \frac{Increase \ in \ log(Likelihood)}{Previous \ Step \ Length}
\end{equation}

One then examines the $Improvement$ score, and alters the length of the
gradient step accordingly.  If we are near the maximum then we would expect
the $Improvement$ score to be rather low, indicating that it is difficult to
improve the likelihood by an appreciable amount.  However, if we are nowhere
near the maximum then we might expect the value of the $Improvement$ to be
rather larger, indicating that it is fairly easy to improve the likelihood
and that a longer step length should be taken.

Of course this method is all rather approximate.  What we require is a more
exact algorithm which generates the optimum step lengths at a given
position.  That is what the conjugate gradient method describes.
  
  \subsection{Conjugate gradient method}

The conjugate gradient method is a refinement of the variable step length
algorithm of section \ref{vsl}.  After one has calculated a normalised
gradient vector then add on a scalar multiple of this vector to the current
coefficient vector such that the likelihood is maximised under the given
constraints. That is to say, we wish to find a value of $\lambda$ which
maximises the following quantity, $\Psi$;

\begin{equation}
\Psi = L \big( \mathbf{\alpha} + \lambda \nabla \mathbf{L}  \big)
\end{equation}

This method means that the gradient step size is always optimal, and that
therefore we reach the maximum most quickly.  Notice that we are not sure
that this is the {\em global} maximum, and therein lies the failing of this
method, and indeed the variable step length method.  It is often rather
straightforward to find a {\em local} maximum of the likelihood function,
but ensuring that this is a {\em global} maximum requires more
consideration. However, this method is an improvement on the method outlined
in section \ref{vsl}, as the following hypothetical illustration
demonstrates.

\begin{figure}
\vspace*{6.5cm}
\includegraphics{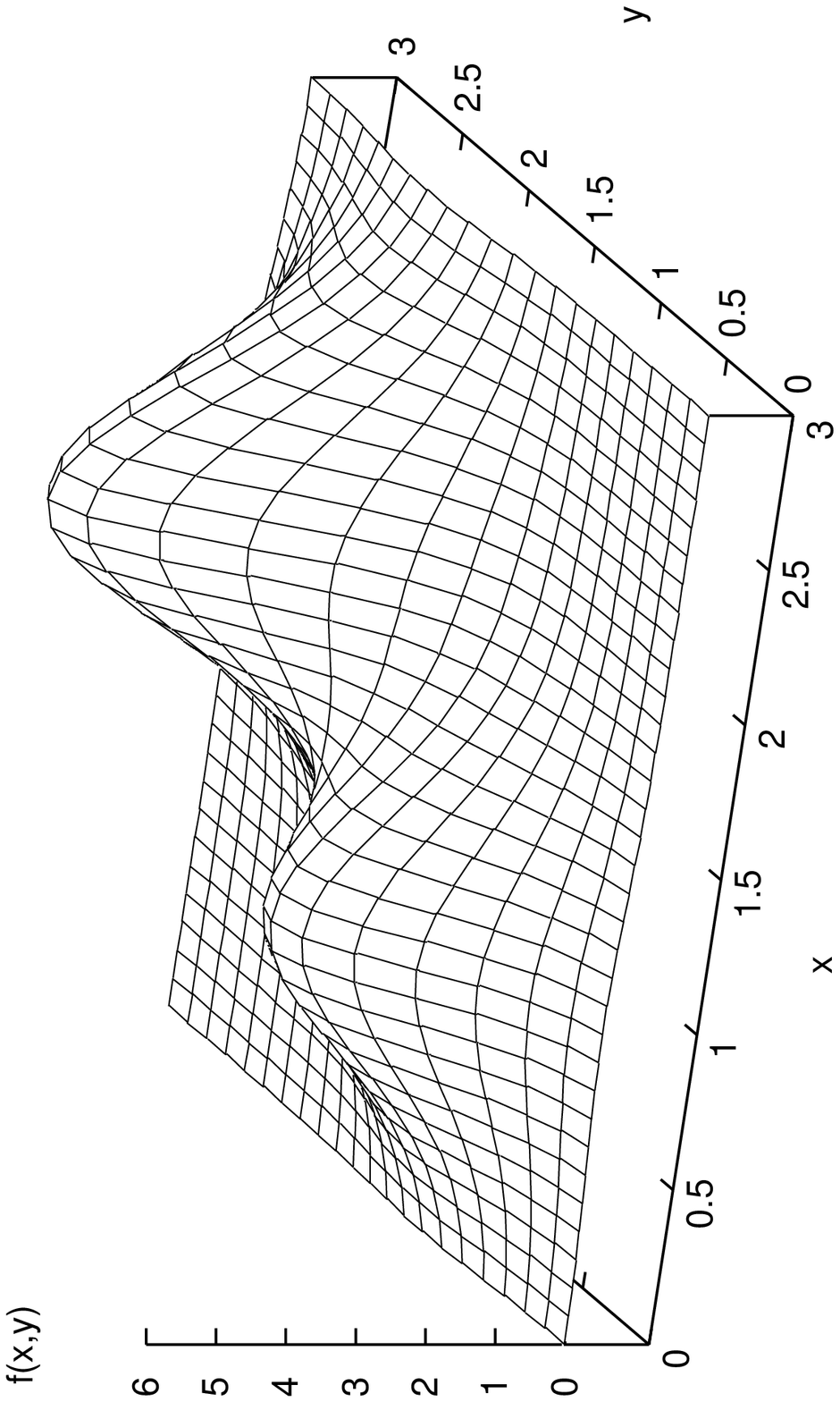}
\vspace*{6.5cm}
\includegraphics{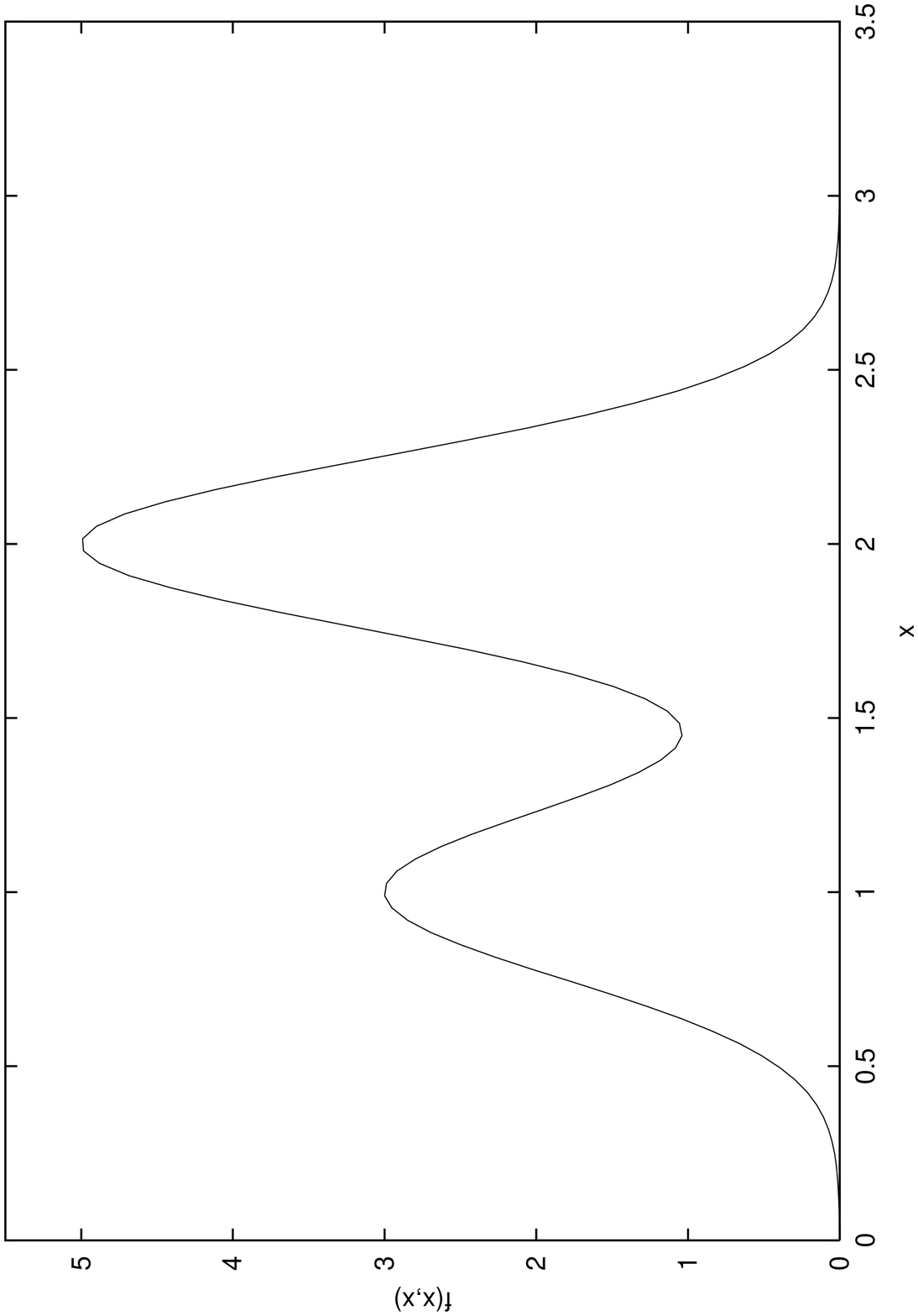}
\caption{Two Gaussian peaks in the (x,y) plane.  One is located at (1,1)
with magnitude 3 units.  The larger is located at (2,2) with magnitude 5
units. Underneath is a cross-section along the line $y=x$.}
\label{hypothetical1}
\end{figure}

Consider we are maximising a function $f$ of two variables, x and y, over the
positive domain.  This function $f$ has two local maxima at $(1,1)$ and
$(2,2)$.  The latter is a slightly larger maximum, and is therefore the
global maximum.  The maximisation procedure starts at the origin.  This
situation is illustrated in figure (\ref{hypothetical1}).

Using the gradient walker technique, it is easy to see that the actual
maximum found depends to a large extent on the choice of gradient steps.  If
the gradient step takes the current coordinates nearer to the slightly
smaller peak then the chances are that will be the one that is found to be
the maximum.  It is clear that a short step length will preferentially climb
the nearer peak first, eventually reaching the local maximum without even
considering the second, slightly larger but more distant maximum.

However, with a long step length, one can envisage a situation where both
peaks are ``overshot''.  One could even end up with an oscillatory set of
solutions in turn overshooting in the positive direction, then overshooting
back in the negative direction.

It is clear that the variable step length method has many serious problems,
which are solved by using the conjugate gradient method.  In this example,
the conjugate gradient method would immediately find the global maximum from
the initial position, as the gradient is in the direction $(1,1)$, which
passes through both peaks.

\begin{figure}
\vspace*{6.5cm}
\includegraphics{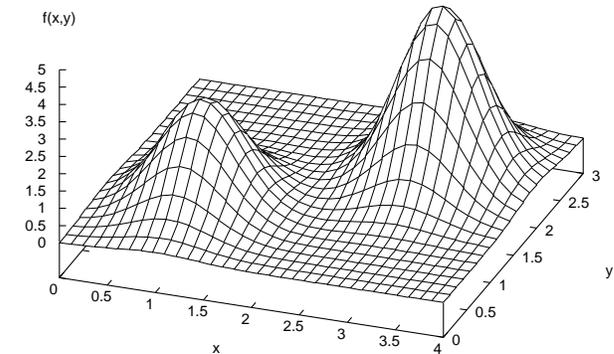}
\vspace*{6.5cm}
\includegraphics{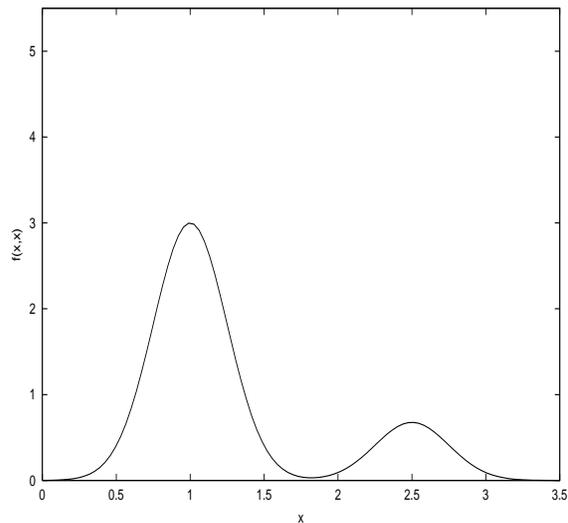}
\caption{In this diagram, the larger peak of magnitude 5 units has been
moved to position (3,2). c.f. figure \ref{hypothetical1}.  Underneath is a
cross section along the line $y=x$.}
\label{hypothetical2}
\end{figure}

However, one could imagine a more likely situation where the second, larger
peak does not lie on the same line as the smaller peak.  Imagine the above
situation with the second peak displaced from $(2,2)$ to $(3,2)$, but with
the same magnitude as in figure \ref{hypothetical1}.  This situation is
shown in figure \ref{hypothetical2}.  The gradient at the origin would still
point approximately towards the peak at $(1,1)$, but would miss the larger
peak completely.  Even the conjugate gradient method would fail here, and
return the smaller local maximum instead of the larger, global maximum.

A new algorithm is required which is more efficient at finding global
maxima, and furthermore deals more effectively with larger dimensionality.
An algorithm ideal for such an application is ``simulated annealing''.

  \subsection{Simulated Annealing}

The method of simulated annealing draws from the realm of materials science,
but only in concept (Kirkpatrick, Gelatt \& Vecchi, 1983a). It is designed
to simulate stochastically the cooling of a physical system in a
sufficiently general sense for it to be applicable to many other
optimisation problems where the global minimum or maximum of a function is
required.

Simulated annealing requires three components.  Firstly, a cost function,
related loosely to the Hamiltonian in a real physical system.  This is the
function which is to be extremised, in this case the logarithmic likelihood
function of the coefficients, $\alpha$.  Secondly, we require some method of
generating random steps within the constrained region in which the cost
function is to be extremised.  Finally, some `temperature' must be defined,
which affects the degree to which random steps are accepted based on the
variation of the cost function.  This temperature obviously corresponds to
no real quantity, it is simply a device used for algorithmic purposes.

This `temperature' parameter is decreased over time either linearly, or in
some more complicated manner, see section \ref{tempdec}.  Random steps are
generated, and are accepted depending on the value of the Hamiltonian or
cost function at the new point relative to the previous point. In most
cases, this cost function is to be minimised, but with our problem the cost
function is the logarithmic likelihood function, which we wish to be
maximised.  Any step which increases the likelihood is automatically
accepted.  Furthermore, any step which decreases the likelihood is accepted
with a certain probability given by;

\begin{equation}
P(step) = exp \bigg( \frac{\Delta log(L)}{T} \bigg)
\end{equation}

Here, $T$ is the `temperature' value described above. For all $\Delta L <
0$, this acceptance probability is less than unity.  The larger the drop in
the likelihood, the smaller the chances of accepting the step.  However, at
higher temperatures, `backward' steps are more likely.  This stops the
optimisation from rapidly centering on the nearest local maximum without
considering the global situation.  Clearly, if a large maximum is found at
any point then the chances of leaving it are smaller because most steps will
lead to significant drops in the likelihood.

As the temperature drops closer to zero, detrimental steps become
increasingly less likely, and the optimisation algorithm begins to centre on
the current maximum, which is expected to be the global maximum.

Clearly some care needs to be taken in the correct implementation of this
algorithm as there are so many different values to fine-tune.  The main two
areas where some thought is required are in the generation of random steps,
and in the temperature decrement function.

    \subsubsection{Choosing random steps}
    \label{randomsteps}

Clearly, with a dimensionality of several hundred, it is important that a
sensible random step is chosen such that the probability of moving in a
useful direction is maximised.  However, one cannot simply bias the step
significantly by the local gradient as this would then remove the advantages
of this method in avoiding local maxima which were not the largest globally.

After some experimentation, we discovered that the most efficient way of
altering the coefficients was to perturb each of the coefficients by a value
proportional to the coefficient itself, at the 10 percent level.  That is to
say that for each coefficient, $\alpha_i$, the perturbation obeyed $-0.1
\alpha_i < \delta_i < 0.1 \alpha_i$.

One further optimisation which could be performed on the step generation
algorithm is simply to add some small percentage of the current local
gradient on to the step to bias the solution towards high likelihood areas.
An analogy of the simulated annealing method is to imagine a ball bouncing
across a 2D surface, where at each point a small percentage of its internal
energy is lost.  In the minimisation algorithm, we require that the ball
eventually finds itself in the deepest minimum in the range of the
variables.  In our scheme, we require the maxima to be found, but the
concept of balls falling into holes is much more straightforward to
visualise, and the algorithm is absolutely identical except for a sign
change.

In the above example, we can cause the ``ball'' to jump preferentially
towards the areas we are interested in by adding on a small fraction of the
local gradient.  Considering a minimisation problem, that effectively means
that the ball will preferentially bounce downhill, as one would expect.  One
modification to the step generation algorithm therefore would be to add on a
certain fraction, $\beta$, of the local gradient, where $\beta$ is a
function of the system temperature, $T$.  That is to say that at high
temperature the `bouncing' is largely random, whereas as the temperature
falls it becomes increasingly biased towards moving towards the minimum
points.  Again, this requires considerable fine-tuning.

For our work, we considered a simple inverse function to generate the
dependence of $\beta$ on the temperature, $T$.  That is, $\beta \propto
\frac{1}{T}$.  As $T$ decreases, the fraction of the gradient added on
increases towards very large values.  In theory we encounter problems as $T$
tends towards zero, but in practice we truncate the search before $T$
actually reaches zero, so these floating point errors are avoided.

In order to maintain the normalisation constraints on the gradient step,
that is to say the total step length remaining roughly constant over time,
the true form of the perturbations becomes the following;

\begin{equation}
\alpha_i \Rightarrow \alpha_i \big( 1 \ + \ \frac{\beta \nabla_i log(L) +
R[-1,1]} {10 (\beta + 1)} \big)
\label{perturb}
\end{equation}

This assumes that the gradient is normalised to unit length.  In this
notation, $R[-1,1]$ represents a random number uniformly distributed in the
range -1 to +1.

Now that we have established a realistic step generation function, it is
necessary to determine the way in which the temperature decreases over time,
as well as the initial value of the temperature.

    \subsubsection{Temperature decrement function}
    \label{tempdec}

In this work we consider only simple temperature decrements of the form $T
\Rightarrow a T - b$.  That is to say a geometric part and an arithmetic
part.  $a$ and $b$ are free parameters here, as is the initial value of the
temperature, $T_0$.

There are no hard-and-fast rules for determining these quantities, so they
were all varied until an optimal value appropriate to the implementation at
hand was discovered. Varying the value of $T_0$ alters the amount to which
random steps are important near the beginning of the optimisation procedure.
The larger the value of $T_0$, the more likely the particle is initially to
take a totally random step to a position of lower likelihood.  Whereas this
is rather useful if one does not have a good initial guess of the global
maximum, this is not the case here.  We can efficiently start with a
reasonably low value of $T_0$, and then alter it slightly depending on the
values of $a$ and $b$ and the total number of steps required.

As for the values of $a$ and $b$, a little more experimentation was
required.  We initially attempted to find optimal solutions with either one
removed in turn.  Firstly, with $a=1$ and $b>0$, that is a purely linear
decrease in $T$, we found that the solutions were not optimal for several
reasons.  Most importantly, the majority of the optimisation happens at low
T when the method simply requires to centre in on the final maximum point.
With a linear decrease, the time spent at low temperature is exactly the
same as the time spent at high temperature for a unit temperature interval.
That means that there is no bias towards locating an accurate maximum
against randomly wandering about parameter space.

However, if we set $b$ to zero and we are left with a purely geometric
decrement function, then we spend far too long at low temperatures.  Indeed,
a low-$T$ cutoff must be introduced in order to prevent the search continuing
for ever.  The benefits in spending so much time at low temperature are
unclear, and probably negligible.  By experimenting with the value of $a$
between 0.95 and 1.00, it was possible to test a reasonable segment of
parameter space in order to test the time it took for a maximum to be found.

We combined the two parameters together on an example dataset, a single
delta function population at 10Gyr and $[Fe/H] = -1.0$. In table (1) we list
the number of steps required and the final maximum likelihood obtained for
different combinations of $a$ and $b$ using a value for the initial
`temperature' of $T_0 = 8$.  Values of $a$ run across the top and $b *
10^{4}$ down the left hand side.  For testing purposes, a uniform
distribution was chosen as a first guess in all cases.

Note that the logarithmic probability values are often greater than zero. As
previously stated, these are only {\em relative} values so the absolute
values are not important, just their differences.  Of course, there is also
a certain amount of random variability in this too, but repeated tests on a
few of the settings showed that it was no greater than $\pm 0.2$ in the
maximum likelihood for those close to the seemingly maximal value, and
usually much less. For those coefficients which didn't get near to this
value then the variation was significantly more, but never brought the
likelihood high enough to be worth considering.

\begin{table*}
\begin{tabular}{|l|l|l|l|l|l|l|l|}
\hline
  & a=  & 0.995     & 0.996      & 0.997      & 0.998      & 0.999      & 1 \\
\hline
  & 0 &211.1/1793 & 211.5/2243 & 211.5/2992 & 211.7/4490 & 211.6/8993 & - \\
  & 1 &162.3/1187 & 205.5/1431 & 211.0/1816 & 211.4/2529 & 211.4/4383 & 211.6/79955 \\
$b*10^4$ = & 2 &-176.7/1054& 158.6/1263 & 209.9/1592 & 211.2/2191 & 211.2/3707 & 211.6/40014\\
  & 3 &-360.4/975 & 17.4/1165  & 207.6/1460 & 210.9/1993 & 211.1/3316 & 211.4/26668\\
  & 4 &-426.2/919 & -91.0/1094 & 202.7/1366 & 210.9/1853 & 211.1/3041 & 211.5/19996\\
  & 5 &-480.3/875 &-226.5/1040 & 200.0/1294 & 210.4/1745 & 211.1/2830 & 211.7/15996\\
\hline
\end{tabular}
\caption
[Optimising the Simulated Annealing T-Decrement Function]
{The effect of changing the values of the coefficients a and b in
equation $T \Rightarrow a T - b$.  Values given are final likelihood/
number of steps required.  Optimal values seem to be around a balanced
medium of these two parameters, such as $a=0.998$, $b=0.0002$. }
\end{table*}

Clearly, it is important to spend some time at low temperature values
optimising the coefficients towards a local maximum.  This is shown by the
clear trend in likelihood with increasing $a$.  However, the number of steps
also increases with increasing $a$, so a balance must be found.

  \subsection{First-guess isochrone coefficients}
  \label{firstguess}

The initial guess for the isochrone coefficients is calculated by
considering the number of stars which are best fit by each isochrone.  The
expectation is that a good first guess will prevent the optimisation
procedure from falling into incorrect local maxima of the probability
function instead of the global maximum.  The likelihood calculation should
also proceed more quickly if many of the original coefficients can also be
ignored right from the start, as this would allow us to sum over fewer
variables.

However, this method introduces several important dangers.  Firstly, one
must always exercise caution when adding prior data to any such optimisation
problem.  If the prior knowledge is either inaccurate or simply misleading
then there is a possibility that it might bias the output of the test.
Secondly, it is important to test that the general form of the recovered
metallicity distribution is not actually a strong function of the chosen
first-guess coefficient distribution.

To test these results, we generated an artificial dataset composed of 500
stars generated from a triple input population of thin metallicity spikes at
metallicity values of $[Fe/H] = -1.0, -0.8$ and $-0.6$.  We then carried out
the optimisation procedure using the best first-guess coefficients to test
the recovery quality.  We also tested the recovery using a flat first-guess
with all coefficients set to the same value.  Thirdly, we tested using
completely random initial coefficients using a standard random number
generator with an unbiased, flat probability distribution.

\begin{figure}
\vspace*{7.2cm}
\includegraphics{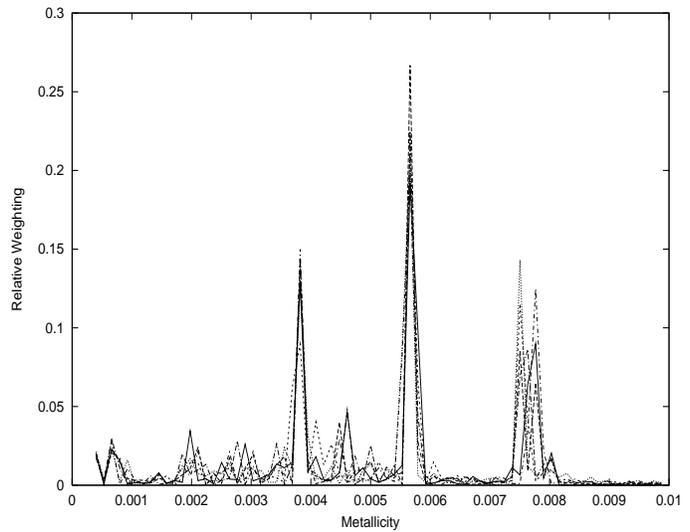}
\caption{Testing the results gained as a function of the initial coefficient
distribution using a best-guess coefficient set (solid line), a flat
distribution (dot-dashed line) and completely random metallicity distributions
(other lines).  Clearly the initial coefficient distribution does not bias
the output strongly.}
\label{testprior}
\end{figure}

Plotted in figure \ref{testprior} are the recovered metallicity
distributions for the above populations.  There are six distributions plotted,
and it is clear that none of these differs from the original recovered
distribution by more than a small amount.  This is reassuring, and
demonstrates that the recovered metallicity distribution is not a strong
function of the input distribution, though the assumption of a realistic
first-guess coefficient distribution does slightly improve the overall fit.

Rather encouragingly, the third metallicity spike appears double in the
diagram, with the first-coefficient guess slightly biasing the output
result.  However, the adopted method, that is the method where the initial
coefficients are assigned best-guess values, produces by far the most
accurate fit to this third population.  In general, the form of the fitting
curves is unchanged regardless of initial coefficient distribution.

All three central values seem to be slightly underestimated in this
particular test.  This is an interesting artefact of the photometric errors.
Broadening the narrow-abundance populations using Poisson errors meant that
stars were spread out over a wider metallicity range.  In this situation,
our code will tend to underestimate the metallicity value because the
isochrones towards lower metallicity are more closely packed. This means
that a slight reduction in the peak metallicity value will fit a larger
number of isochrones more accurately.  Ideally we should try to space the
isochrones evenly in colour-magnitude separation rather than in metallicity,
but this is not possible in practice because the isochrones are not
parallel.

  \subsection{Monte-Carlo enhancements}

One way to solve this problem is to introduce the idea of Monte-Carlo steps
after the temperature drops to a certain low threshold.  In a sense, this
means continuing the simulated annealing algorithm, but with a zero
temperature.  Random steps are taken and accepted only if they improve the
likelihood.  In the previous analogy, it is like allowing the ``ball'' to
roll down the ``holes'' towards an optimum value.

To implement this method, we used similar routines to the simulated annealing
approach.  Firstly, a subset of the coefficients were selected, and then
perturbed by a certain random percentage, as above.  A new likelihood was
then calculated, and if it was an improvement then the step was taken.  If
it was not an improvement then the new coefficients were rejected.

This algorithm was continued until a certain number of quiescent
perturbations had been performed.  We defined ``quiescent'' as perturbations
which did not improve the likelihood above a certain small value.  This
stopped the Monte-Carlo optimisation routine from continuing ad infinitum
with infinitesimally small improvements which had no real effect on the
overall metallicity distribution recovered.

We also used a slightly smaller step length in this algorithm, as we were
expecting the coefficients to already be reasonably near the global
maximum, so only small perturbations were required.  Clearly we require
a reasonably accurate guess to the maximum from the annealing method or
the Monte-Carlo steps could potentially continue for an unreasonably long
time.

In addition, after every perturbation step the coefficients perturbed were
recorded, together with the direction in which they were perturbed.  If a
perturbation improved the likelihood then the same coefficients were
reconsidered with the same perturbation directions.  This continued until
the likelihood no longer improved.

This method allows a much more accurate determination of the true maximum
point to be established.  The overall effect was never to truly change the
metallicity distribution, so this method was only used when the most
accurate answer possible was required.  Clearly there is a certain level
after which this kind of further optimisation is irrelevant.  The errors
inherent in the optimisation procedure itself together with the method of
assigning initial probabilities have a far greater effect.  However, the
method is certainly generally applicable to problems where greater
accuracy is both possible, and required.

\begin{figure*}
\vspace*{6.5cm}
\parbox{7cm}{
\begin{center}
\includegraphics{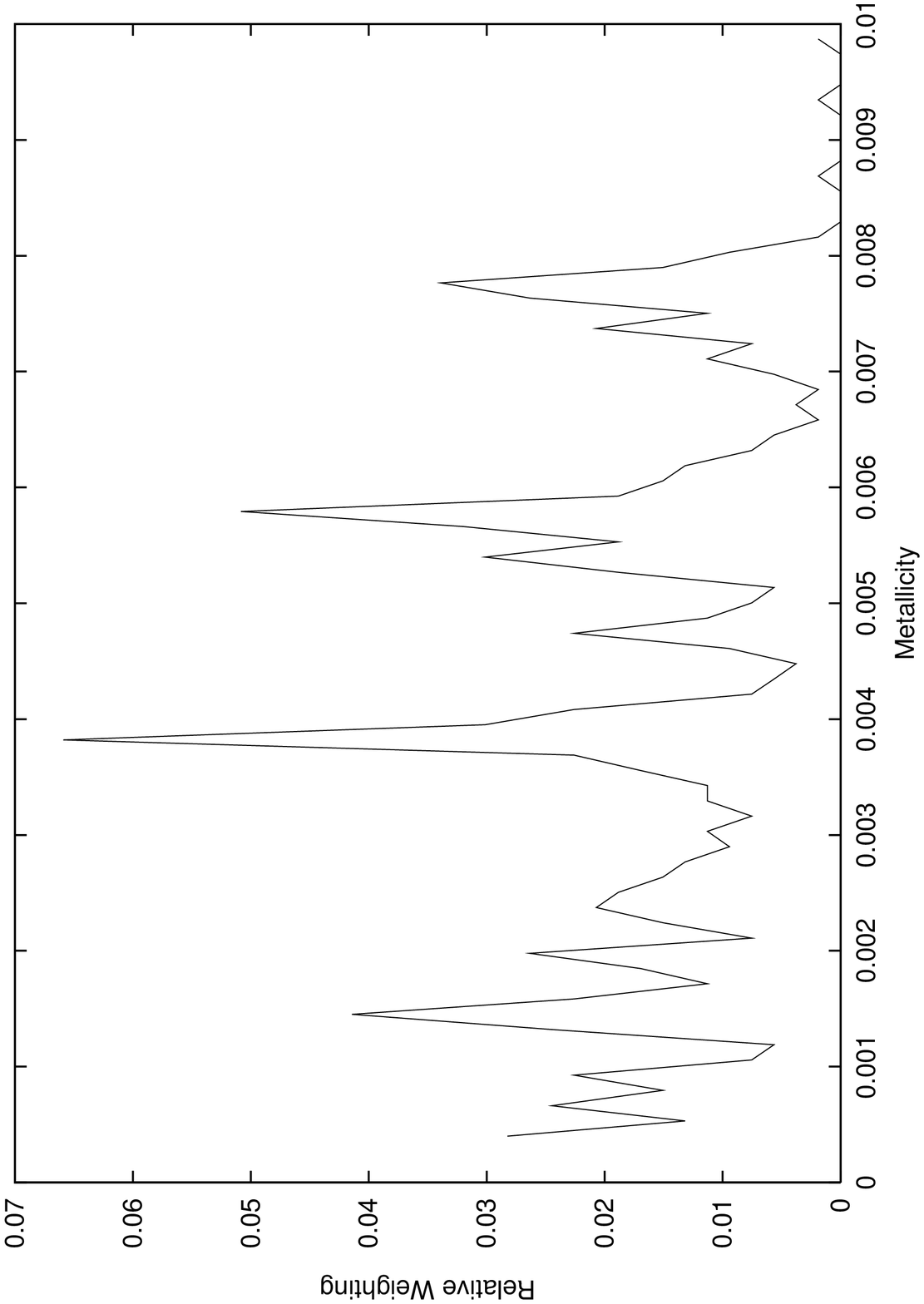}
\end{center}}
\hspace{1cm}
\parbox{7cm}{
\begin{center}
\includegraphics{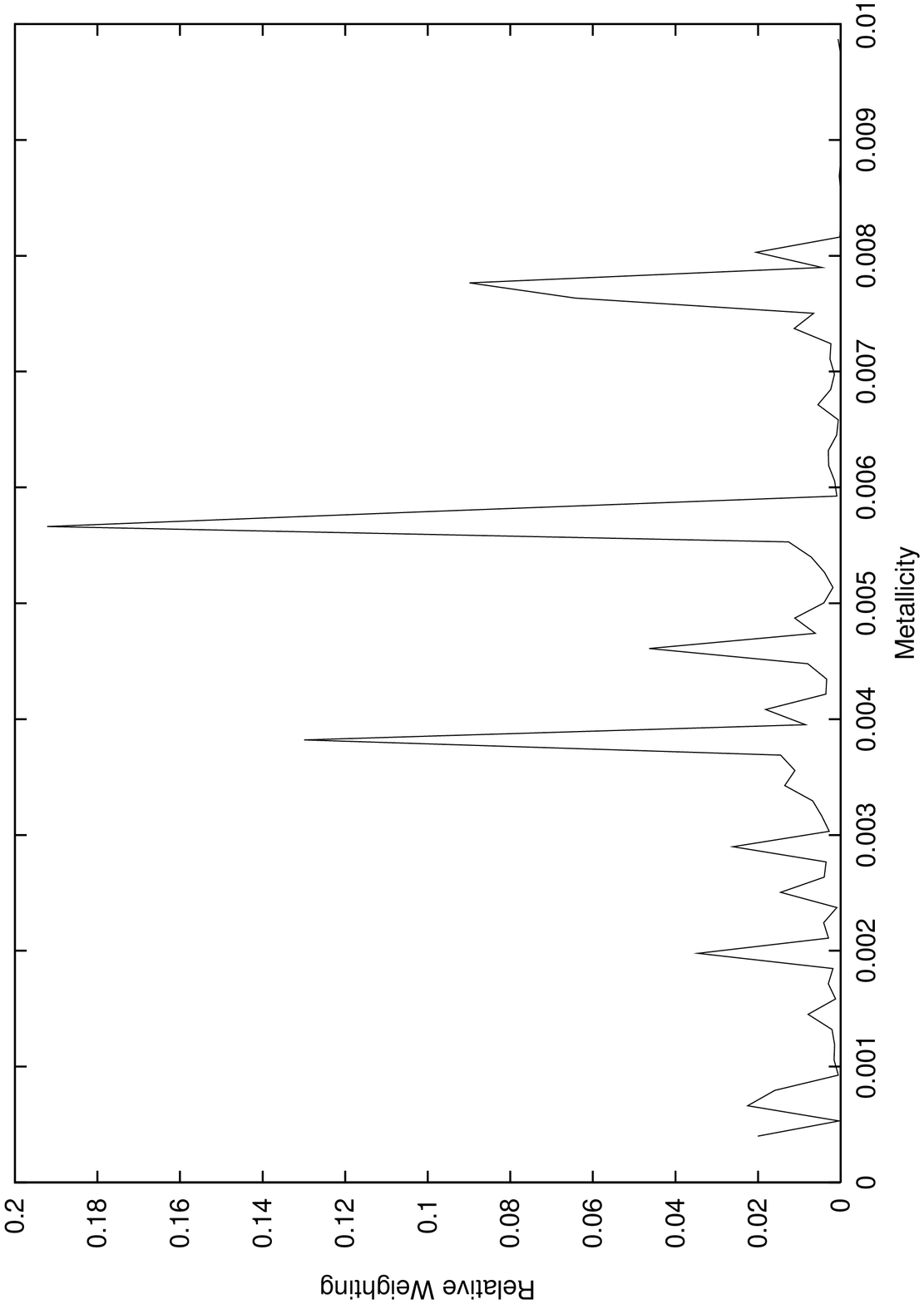}
\end{center}}
\caption{Left: The first guess coefficient values for the sum of three equal
size delta populations at metallicities of $[Fe/H] = -1.0, -0.8$ and $-0.6$.
Poisson errors and 3 per cent photometric errors were introduced into the
simulated dataset.
Right: The final optimised version after simulated annealing and Monte-Carlo
optimisations.  Residual errors are caused by the Horizontal Branch, which
is notoriously difficult to fit.  By fitting only the RGB, the quality of
fit is significantly improved.  See the next paper in this series for a more
rigorous investigation.}
\label{appendix1}
\end{figure*}

\begin{figure*}
\vspace*{6.5cm}
\parbox{7cm}{
\begin{center}
\includegraphics{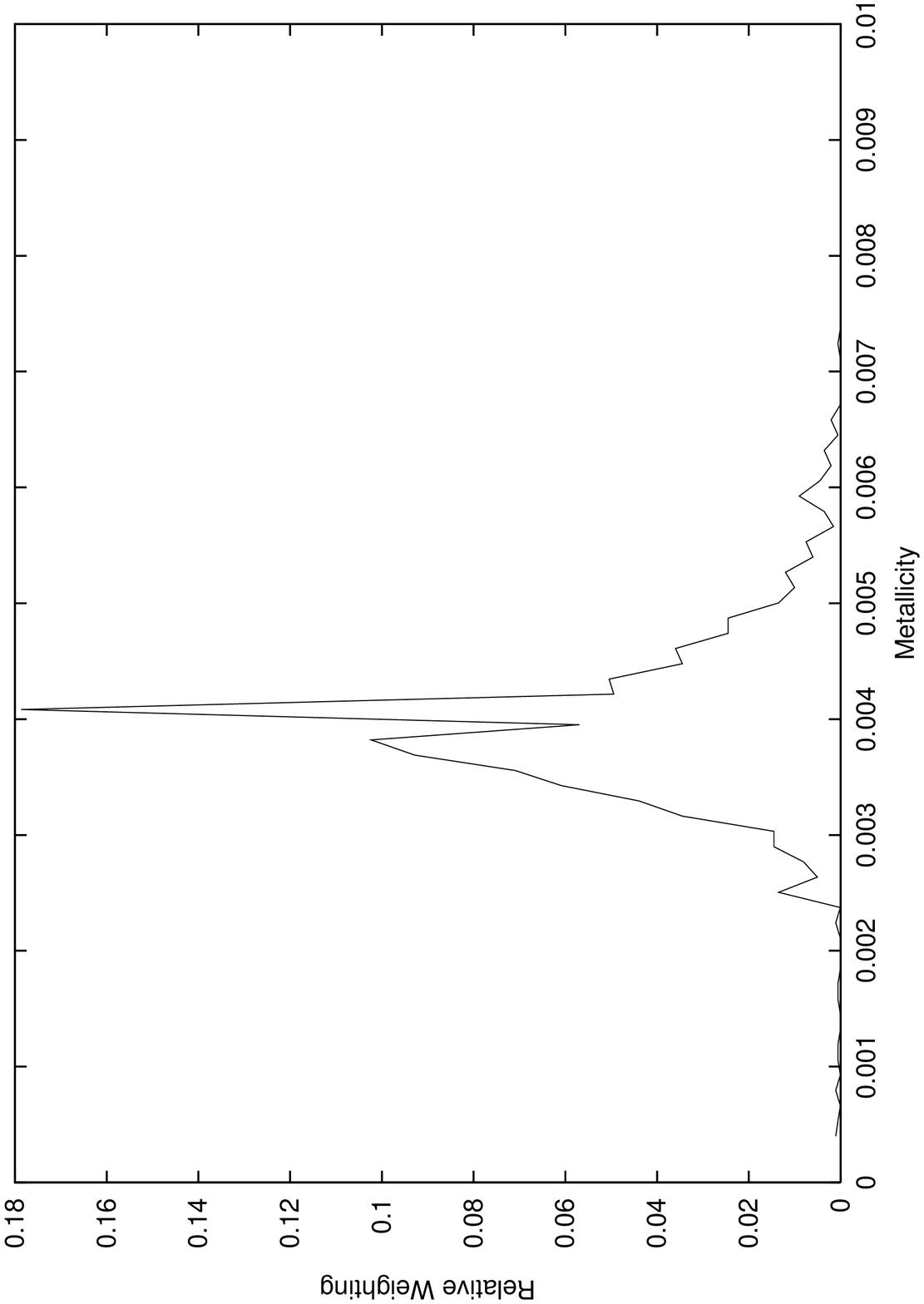}
\end{center}}
\hspace{1cm}
\parbox{7cm}{
\begin{center}
\includegraphics{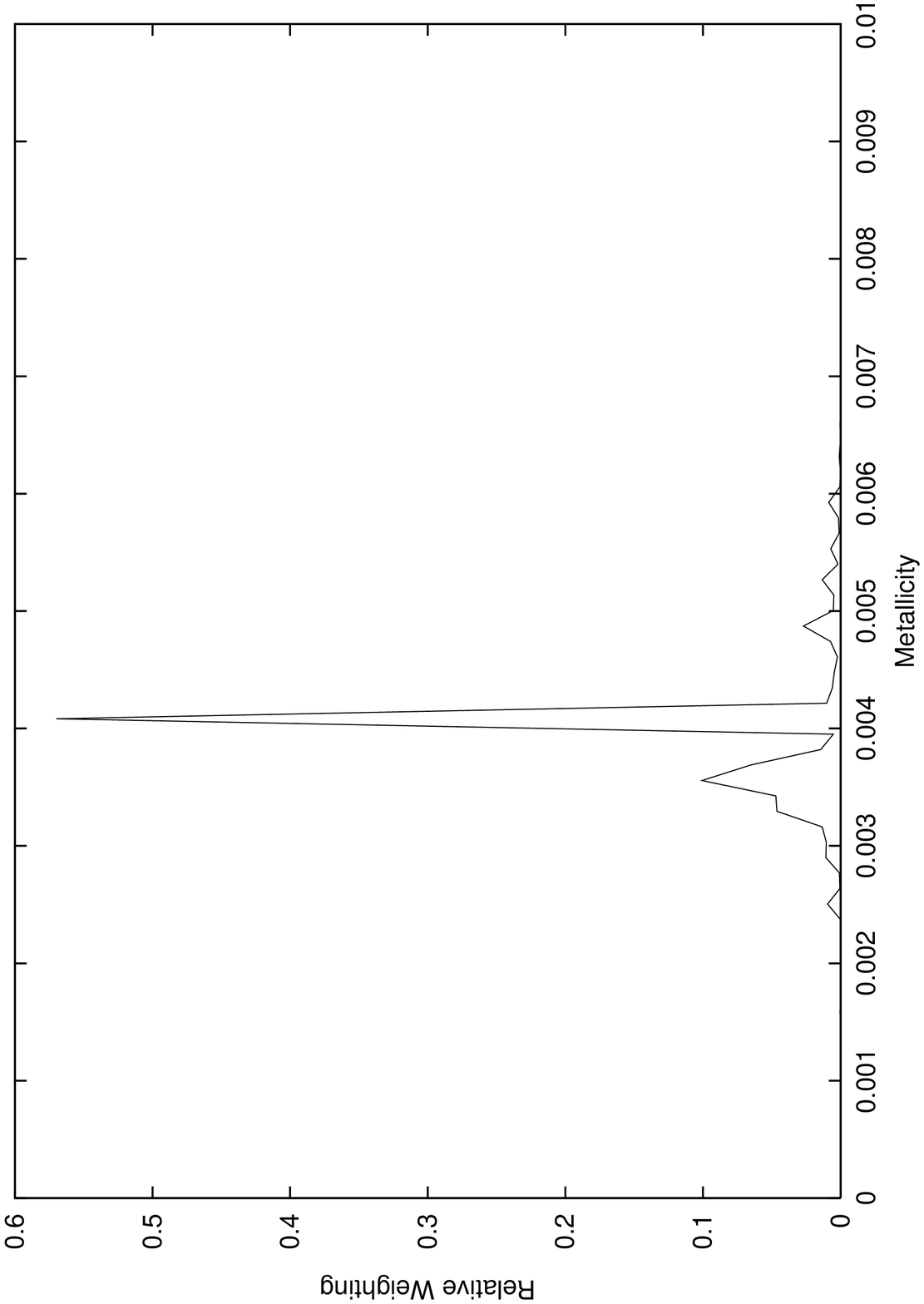}
\end{center}}
\caption{Calculated best fit coefficient values for a non-coeval population
centred on 10Gyr and a metallicity of $[Fe/H] = -1.0$.  We added a narrow age
spread of $\pm 0.3$ Gyr into the initial data in order to test the degree to
which this affected the recovery of the central metallicity value. Poisson
errors and 3 per cent photometric errors were introduced into the simulated CMD.
Left: First guess before optimisation. Right: The final version after
simulated annealing and Monte-Carlo optimisations.}
\label{appendix2}
\end{figure*}

In figures \ref{appendix1} and \ref{appendix2} we demonstrate a small
selection of results generated by the metallicity distribution fitting
programme, $FITCOEVAL$.  We have tested this programme on the artificial
datasets generated by the code described above. Diagram captions explain in
greater detail the fitting methods employed.

\section{Conclusion}

We have presented appropriate tools and methods for generating and analysing
colour magnitude diagrams for old giant-branch stellar populations.  Our
subsequent work has allowed us to test the accuracy of these methods, and to
develop new algorithms to introduce Bayesian prior knowledge into the
fitting procedure.

Using the tools outlined in this paper we can now do the following;

\begin{enumerate}
\item Create artificial colour-magnitude diagrams, paying attention to the
many different sources of error inherent in conventional observational
techniques.
\item Compare artificial colour magnitude diagrams to those obtained from
observations using any of a number of methods, including that outlined by
Harris \& Zaritsky (2001).
\item Fit a maximum likelihood metallicity distribution to old RGB stellar
populations using an interpolated grid of theoretical isochrones.
\end{enumerate}

The second paper of this series (to be published soon) discusses the
limitations and systematic errors involved in RGB isochrone fitting methods.
We treat the handling of errors both in the computational methods employed,
and also in the isochrone models themselves.  We derive limits to which
isochrone fitting can help us to obtain information about the RGB
populations of nearby galaxies.

\section{Acknowledgments}

C.M.F. would like to extend his thanks to Jan Kleyna for help in the
preparation of the code required for this project, and also to Richard de
Grijs for many interesting discussions on these topics and valuable comments
on early drafts of the paper.


\newpage

\label{lastpage}
\end{document}